\documentclass[prl,noshowpacs,english,superscriptaddress,twocolumn]{revtex4}
\usepackage{amsmath,amssymb,amsfonts,mathrsfs}
\usepackage{amsthm}
\usepackage{CJK}
\usepackage{bm}
\usepackage{babel}
\usepackage{graphicx}
\allowdisplaybreaks
\usepackage{braket}
\usepackage[breaklinks]{hyperref}
\hypersetup{colorlinks=true, linkcolor=blue, citecolor=blue, filecolor=blue, urlcolor=blue}

\begin{document}

\title{Room-Temperature Ferromagnetic Topological Phase in CrO$_{2}$: From Tripe Fermions to Weyl Fermions}

\author{R. Wang}
\affiliation{Department of Physics, Southern University of Science and Technology,
Shenzhen 518055, P. R. China}
\affiliation{Institute for Structure and Function $\&$
Department of physics, Chongqing University, Chongqing 400044, P. R. China}
\author{Y. J. Jin}
\affiliation{Department of Physics, Southern University of Science and Technology,
Shenzhen 518055, P. R. China}
\author{L. Duan}
\affiliation{Institute of Physics, Chinese Academy of Sciences, Beijing 100190, P. R. China}
\author{J. Z. Zhao}
\affiliation{Department of Physics, Southern University of Science and Technology, Shenzhen 518055, P. R. China}
\author{H. Xu}
\email[]{xuh@sustc.edu.cn}
\affiliation{Department of Physics, Southern University of Science and Technology, Shenzhen 518055, P. R. China}
\affiliation{Institute for Quantum Science and Engineering, Southern University of Science and Technology, Shenzhen 518055, P. R. China}
\author{C. Q. Jin}
\email[]{jin@iphy.ac.cn}
\affiliation{Institute of Physics, Chinese Academy of Sciences, Beijing 100190, P. R. China}

\begin{abstract}
Ferromagnetic topological semimetals due to their band topology co-existing with intrinsic magnetization exerted important influences on early study of topological fermions. However, they have not been observed in experiments up to now. In this work, we propose that rutile CrO$_{2}$, a widely used half-metallic ferromagnetic material in magnetic recording taps, exhibits unexpected ferromagnetic topological features. Using first-principles calculations and symmetry analysis, we reveal that rutile CrO$_2$ hosts the  triple nodal points in the absence of spin orbital coupling (SOC). By taking into account of SOC, each triple nodal point splits into two Weyl points, which are located on the magnetic axis with four-fold rotational symmetry. Notably, the Fermi arcs and accompanying quasiparticle interference patterns are clearly visible, which facilitate experimental observations. In addition, we find that another experimentally synthesized  CrO$_{2}$ in CaCl$_2$ structure, also hosts the topologically nontrivial ferromagnetic phase. Our findings substantially advance the experimental realization of ferromagnetic topological semimetals. More importantly, room-temperature time-reversal-breaking Weyl fermions in CrO$_{2}$ may result in promising applications related to the topological phenomena in industry.
\end{abstract}



\maketitle
Topological semimetals or metals, in which the conduction and valence bands cross each other near Fermi level resulting in topological protected nodal points or nodal lines in momentum space, are an emerging topological phase in a condensed matter system. Accordingly, different fermions with nontrivial band topology have been proposed, i.e., the fourfold-degenerate Dirac fermions \cite{WangPhysRevB.85.195320,ScienceLiu864}, twofold-degenerate Weyl fermions~\cite{Wan2011,PhysRevLett.107.127205,Xu2011}, three-degenerate triple fermions \cite{lvNature2017, ZhuPhysRevX.6.031003}, and node-line fermions~\cite{PhysRevB.84.235126}, etc. These protected band degeneracies require a striking interplay of symmetry and topology in electronic structures of materials. For instance, Dirac semimetal (DSM) phases are protected by crystal rotational symmetry \cite{WangPhysRevB.85.195320,ScienceLiu864} or nonsymmorphic symmetry \cite{PhysRevLett.116.186402}; Weyl semimetal (WSM) phases require broken either spatial inversion symmetry $\mathcal{I}$ \cite{Weng2015} or time-reversal symmetry $\mathcal{T}$ \cite{Wan2011,Xu2011}; node-line semimetals (NLSMs) require a space-time inversion $\mathcal{I}\mathcal{T}$ with spin-rotation symmetry~\cite{PhysRevB.92.081201, PhysRevLett.116.156402} or a mirror reflection~\cite{PhysRevB.90.205136,PhysRevLett.62.2747}. Distinguished from normal semimetals or metals, the linear quasiparticle excitations in topological semimetals or metals near the crossing points will lead to diverse exotic quantum features. In particular, WSMs which host the unique nodal points called Weyl points (WPs) with definite chirality, acting as a magnetic monopole in momentum space, will exhibit interesting transport phenomena such as surface Fermi arcs, chiral anomaly, and anomalous Hall effects \cite{Wan2011,PhysRevLett.107.127205, Xu2011, PhysRevB.90.155316, PhysRevLett.113.187202}. Recently, WSMs have been intensively studied theoretically \cite{Wan2011,Xu2011, Weng2015, Huang2015,  Ruan2016nc, Ruan2016prl, Autes2016, Wang2016prl2, Wang2016prl1, Xu2017prl, Wang216, Wangrui2017} and have been observed in nonmagnetic TaAs family \cite{Xu2015} and MoTe$_2$ \cite{PhysRevX.6.031021}.

In comparison with nonmagnetic WSMs, ferromagnetic (FM) WSMs with time-reversal symmetry breaking are promising materials for the future spintronic devices due to their nontrivial electronic structures co-existing with intrinsic spontaneous magnetization. However, only a few FM WSMs have been proposed \cite{Xu2011, Wang2016prl1, Wangrui2017}. However, Weyl fermions and accompanying Fermi arcs in FM WSMs have not been successfully verified experimentally. Therefore, searching for reliable FM topological materials that will be verified in experiments is highly desired. In particular, FM topological semimetal materials with Curie temperatures well above room temperature have significant applications.

In this work, we propose that two phases of CrO$_2$, i.e., rutile ($r$-) CrO$_2$ and orthorhombic ($o$-) CrO$_2$ of the CaCl$_2$ type, possess FM Weyl fermions using first-principles calculations and symmetry analysis. $r$-CrO$_2$ is a very well-known half-metallic ferromagnet with a high Curie temperature in the range of 385-400 K \cite{PhysRevLett.50.2024, PhysRevB.75.140406} that is widely used in magnetic recording taps, suggesting that $r$-CrO$_2$ is an ideal candidate for studying FM Weyl fermions in experiments. Crystalline CrO$_2$ undergoes a structural phase transition from a rutile to an orthorhombic structure in the pressure range of $12\sim17$ GPa \cite{PhysRevB.73.144111}. As $o$-CrO$_2$ exhibits similar FM topological properties with $r$-CrO$_2$, we mainly focus on $r$-CrO$_2$ to investigate the topologically nontrivial features of CrO$_2$, and detailed results of $o$-CrO$_2$ are provided in the Supplemental Material (SM) \cite{SM}. Our findings provide realistic room-temperature candidates to realize topological spin manipulations and potential spintronic devices.

\begin{figure}
\setlength{\belowcaptionskip}{-0.4cm}
	\centering
	\includegraphics[scale=0.38]{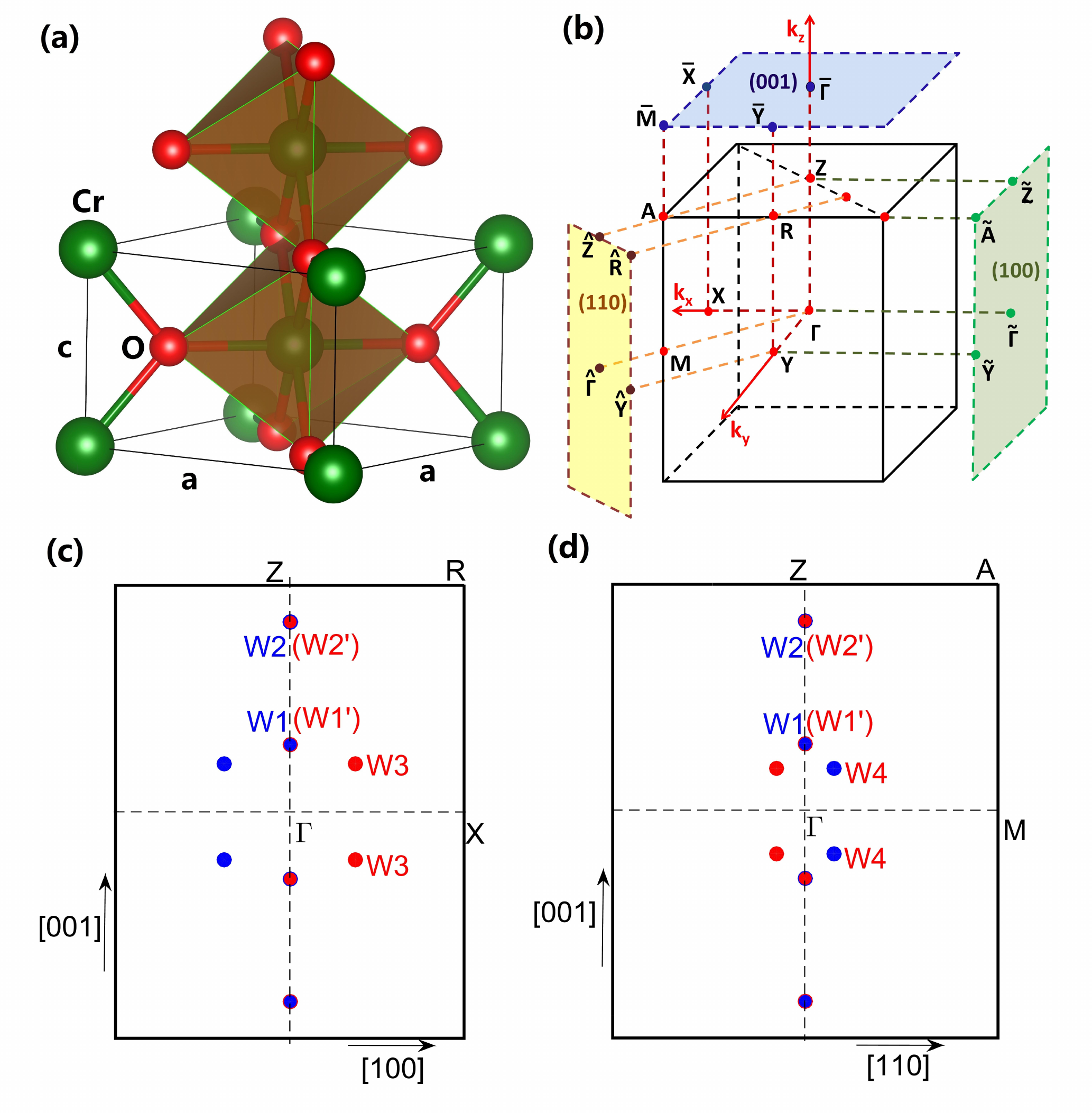}
	\caption{ (a) $r$-CrO$_2$ with space group $P4_2/mnm$ (No. 136). Cr and O atoms are represented by green and red spheres, respectively. (b) The first Brillouin zone (BZ) and the corresponding surface BZ. The distributions of  WPs on (c) $k_x$-$k_z$ and (d)$k_x$=$k_y$ planes. The red and blue dots denote the WPs with Cherm numbers +1 and -1, respectively.
\label{figure1}}
\end{figure}

First-principles calculations are performed using Vienna \textit{ab initio} Simulation Package \cite{Kresse2,Kressecom}. The strong correlation effect of Cr $3d$ electrons is considered by introducing the effective on-site Coulomb energy U=3 eV \cite{Liechtenstein1995}, which gives an appropriate description that is in excellent agreement with previously reported band structures and half-metallic properties of $r$-CrO$_2$ \cite{PhysRevLett.80.4305}. Importantly, FM topological features are robust with U in the range from 1.5 eV to 6.0 eV. More computational details are included in SM \cite{SM}.

\begin{figure}
\setlength{\belowcaptionskip}{-0.4cm}
	\centering
	\includegraphics[scale=0.325]{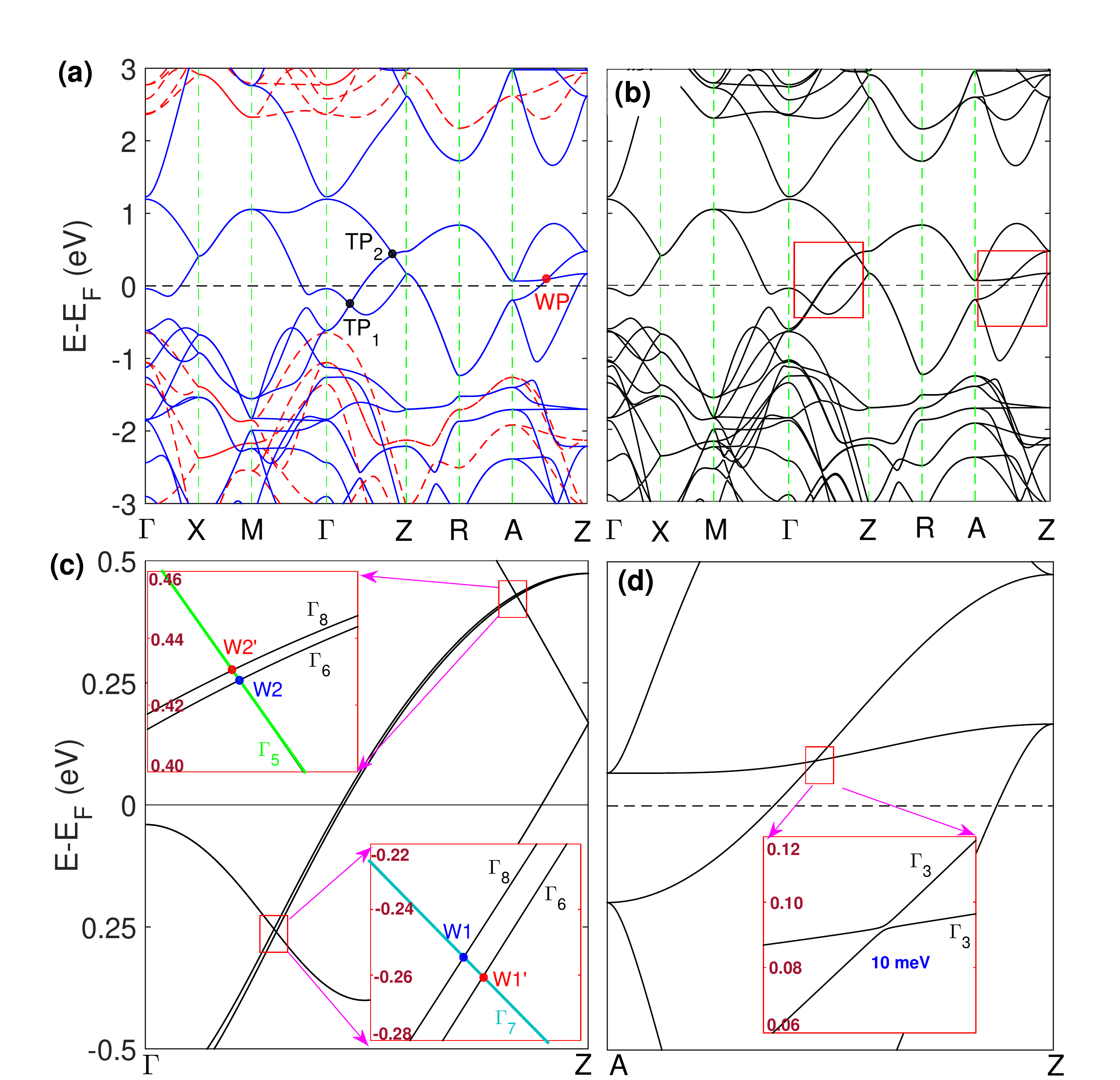}
	\caption{Band structures of $r$-CrO$_2$. (a) Band structure without SOC. The majority and minority spin bands are denoted by the solid and dashed lines. Two triple points (TPs) (i.e., $TP1$ and $TP2$) along the $\Gamma$-Z direction and a type-II WP along the A-Z direction are labelled. (b) Band structure in the presence of SOC with a [001] magnetization direction.  The panels (c) and (d) show the enlarged views along $\Gamma$-Z and A-Z directions in panel (b), respectively. $W1$ ($W1'$) and $W2$ ($W2'$) arise from $TP1$ and $TP2$ splitting, respectively, due to magnetization along [001] direction.
\label{ele-band}}
\end{figure}

In ambient condition, CrO$_2$ crystalizes in a rutile structure with space group $P4_2/mnm$ (No. 136), as shown in Fig. \ref{figure1}(a). The optimized  lattice constants are $a=4.470$ {\AA} and $c=2.973$ {\AA}, which nicely agree with the experimental values $a=4.421$ {\AA} and $c=2.916$ {\AA} \cite{PhysRevB.88.085123}. As shown in Fig. \ref{ele-band}(a), two spin channels behave differently that the spin and orbital freedoms are independent, namely, the majority spin states show metallic properties while the minority spin channel has a insulating band gap. The ground state of $r$-CrO$_2$ is confirmed to be FM with a saturation magnetic moment of $2.00$ $\mu_B$. Our calculations indicate that $r$-CrO$_2$ is completely spin-polarized at the Fermi level, which is in excellent agreement with prior calculations~\cite{Schwarz1986, PhysRevLett.80.4305} and point contact Andreev spectroscopy \cite{PhysRevLett.86.5585}.

In the spin-polarized calculation, neglecting SOC, the point group $D_{4h}$ and spin-rotation symmetry $SU(2)$ coexist in $r$-CrO$_2$. The symmorphic group elements contain inversion $\mathcal{I}$, two-fold rotational symmetry $C_2$ axes oriented along [100], [010], and [110], four-fold rotational symmetry $C_4$ axis along [001](i.e., $z$-axis), and mirror reflection $M_z$. In addition, the space group $P4_2/mnm$ also possesses nonsymmorphic symmetries as $N_x=\{M_x | \mathbf{\tau}\}$ and $N_{4z}=\{C_{4z} | \mathbf{\tau}\}$, where $\mathbf{\tau}=(\frac{1}{2},\frac{1}{2},\frac{1}{2})$ is the translation vector of one-half of a body diagonal. In the absence of SOC, the triple points (TPs) $TP1$ and $TP2$ along the $\Gamma$-Z direction shown in Fig. \ref{ele-band}(a), can be understood by considering symmorphic symmetries $\mathcal{I}$, $M_z$ and $C_{4z}$, and the nonsymmorphic $N_{x}$. Based on symmetry transformation, we have the relation $N_x C_{4z} = C_{2z}T_{(0,-1,0)} (C_{4z}N_x)$ (where $T_{(0,-1,0)}$ is the translation operator acting on the Bloch wave function), i.e., $N_x$ and $C_{4z}$ can not commute each other (see SM \cite{SM}). Every momentum point along the $\Gamma$-Z direction is invariant under $C_{4z}$ and the product $PM_z$. Hence, the little group $C_{4v}$ in the $\Gamma$-Z direction can maintain that a double-fold degenerate band always crosses two single-degenerate bands, resulting in the formation of two TPs denoted as $TP1$ and $TP2$. In addition, two bands belong to different $C_2$ eigenvalues $\pm i$ along the A-Z high-symmetry line cross each other. As a result, a type-II Weyl point (WP) appears, which is protected by $C_2$ rotation symmetry along the [110] axis. Comparing with $r$-CrO$_2$, the lower symmetry of $o$-CrO$_2$ (space group $Pnnm$) can not maintain TPs along $\Gamma$-Z; however, the symmetry-protected WPs existing on this high-symmetry line are found in $o$-CrO$_2$ (see SM \cite{SM}).

Spin-orbital coupling (SOC) only has a insignificant influence on electronic properties of CrO$_2$ due to weak SOC effects of both Cr and O elements. Therefore, the half-metallic ferromagnetism is preserved and the band structure is almost the same in the presence of SOC, as shown in Fig. \ref{ele-band}(b). Although band structures with and without SOC exhibit similar behaviors, they host the different band topology due to symmetry changes caused by the introduction of magnetization.  It is worth to mention that all possible magnetic configurations of $r$-CrO$_2$ are nearly degenerate, implying that magnetization directions can be easily manipulated by applying a magnetic field. Therefore, we only pay attention to the case of the [001] magnetization direction in the main text, and the analysis of [100] magnetization direction is provided in SM \cite{SM}.

\begin{table}
\caption{Nodal points of $r$-CrO$_2$ without and with SOC.  The coordinates in momentum space, Chern numbers, and the energies relative to $E_F$ are listed, respectively.
The coordinates of other WPs are related to the ones listed by the symmetries, $\mathcal{P}$, $C_{4z}$, $\mathcal{T}C_{2x}$, and $\mathcal{T}C_{2y}$.}
  \begin{tabular}{p{1.2 cm}|p{0.9 cm}|*{1}{p{3.5cm}} *{3}{p{1.2cm}} }
  \hline
  \hline
    &Nodal  & \centering Coordinates [$k_x(2\pi/a)$,    &\centering  Chern & $E-E_F$ \\
    &points & \centering $k_y(2\pi/a)$, $k_z(2\pi/c)$] &\centering number & (eV) \\
  \hline
   \centering Without & \centering $TP1$  &\centering  (0,  0,   0.1461) &\centering - &{\centering -0.256} \\
   \centering SOC &\centering $TP2$  &\centering  (0,  0,   0.4188) &\centering -  &{\centering 0.428} \\
   &\centering WP  &\centering  (0.2669    0.2669    0.5) &\centering +1 &{\centering 0.092} \\
  \hline
    &\centering  $W1$  &\centering (0,  0,   0.1451)   &\centering $-1$  &{\centering -0.254} \\
    &\centering  $W1'$ &\centering (0,  0,   0.1477)   &\centering $+1$  &{\centering -0.260}  \\
  With&\centering$W2$ &\centering  (0,  0,   0.4193)   &\centering $-1$  &{\centering  0.427} \\
  SOC&\centering $W2'$&\centering  (0,  0,   0.4181)   &\centering $+1$  &{\centering  0.430} \\
    &\centering  $W3$ &\centering (0.1897,    0,    0.1069)       &\centering $+1$ &{\centering -0.279}  \\
    &\centering  $W4$ &\centering (0.1056,    0.1038,    0.0941)  &\centering $-1$ &{\centering -0.223} \\
  \hline
  \hline
  \end{tabular}
  \label{table}
\end{table}

In the presence of SOC, two spin channels couple together and the spin-rotation symmetry $SU(2)$ is broken. When the magnetization is along the [001] direction, the spatial symmetry is reduced to the magnetic double point group $D_{4h}(C_{4h})$. In this case, $C_2$ rotational symmetries along [100], [010], and [110] axes are broken, while the magnetic symmetry $\mathcal{T}C_2$ is still maintained. The little group of $\mathbf{k}(0, 0, k_z)$ along $\Gamma$-Z is $C_{4}$, which is the subgroup of $D_4$, and the mirror reflection along this direction is broken. As shown in Fig. \ref{ele-band}(c), the intrinsic magnetization along the [001] direction will split the double-fold degenerate band into two single-degenerate bands denoted by $\Gamma_6$ ($e^{-i\frac{\pi}{4}}$) and $\Gamma_{8}$ ($e^{i\frac{3\pi}{4}}$), which cross the bands $\Gamma_5$ ($e^{-i\frac{\pi}{4}}$) and $\Gamma_7$ ($e^{i\frac{5\pi}{4}}$), respectively, where $\Gamma_5$, $\Gamma_6$, $\Gamma_7$, and $\Gamma_8$ are the eigenvalues of four-fold rotation $C_{4z}$. Therefore, each TP splits into two WPs with the opposite chirality, i.e., $TP1$ evolves to WPs $W1$ ($W1'$) and $TP2$ evolves to WPs $W2$ ($W2'$). As shown in Fig. \ref{ele-band}(d), the [001] magnetization destroys the type-II WP along A-Z with a gap opening of $\sim$10meV, as the symmetry $\mathcal{T}C2_{110}$ doesn't allow band crossings on this line. In contrast, the existence of WPs in high-symmetry $k_x$-$k_z$ and $k_y$-$k_z$ planes is allowed by the antiunitary symmetries $\mathcal{T}C_{2y}$ and $\mathcal{T}C_{2x}$, respectively (see SM \cite{SM}). Through carefully checking, we find that WPs $W3$, which relate to $\mathcal{I}$, $\mathcal{T}C_{2x}$, and $\mathcal{T}C_{2y}$, are present in the $k_x$-$k_z$ and $k_y$-$k_z$ planes. In addition, we also find that other WPs $W4$ locate near the $k_x=k_y$ and $k_x=-k_y$ planes, and these WPs related by $\mathcal{I}$ and $C_{4z}$ are generic nodal points, which do not correspond to any little-group symmetry. Hence, the WPs $W4$ are not stable and easily annihilate with each other~\cite{Wang2016prl1}. The precise coordinates, Chern numbers, and energies related to the Fermi level $E_F$ of nodal points are listed in Table \ref{table}. The Chern numbers are calculated using \textit{Z2Pack} package \cite{PhysRevB.95.075146}, in which the evolution of average Wannier charge centers obtained by the Wilson-loop method on a sphere around WPs is employed \cite{Yu2011,Soluyanov2011}.

\begin{figure}
\setlength{\belowcaptionskip}{-0.4cm}
\setlength{\abovecaptionskip}{-0.1cm}
	\centering
	\includegraphics[scale=0.42]{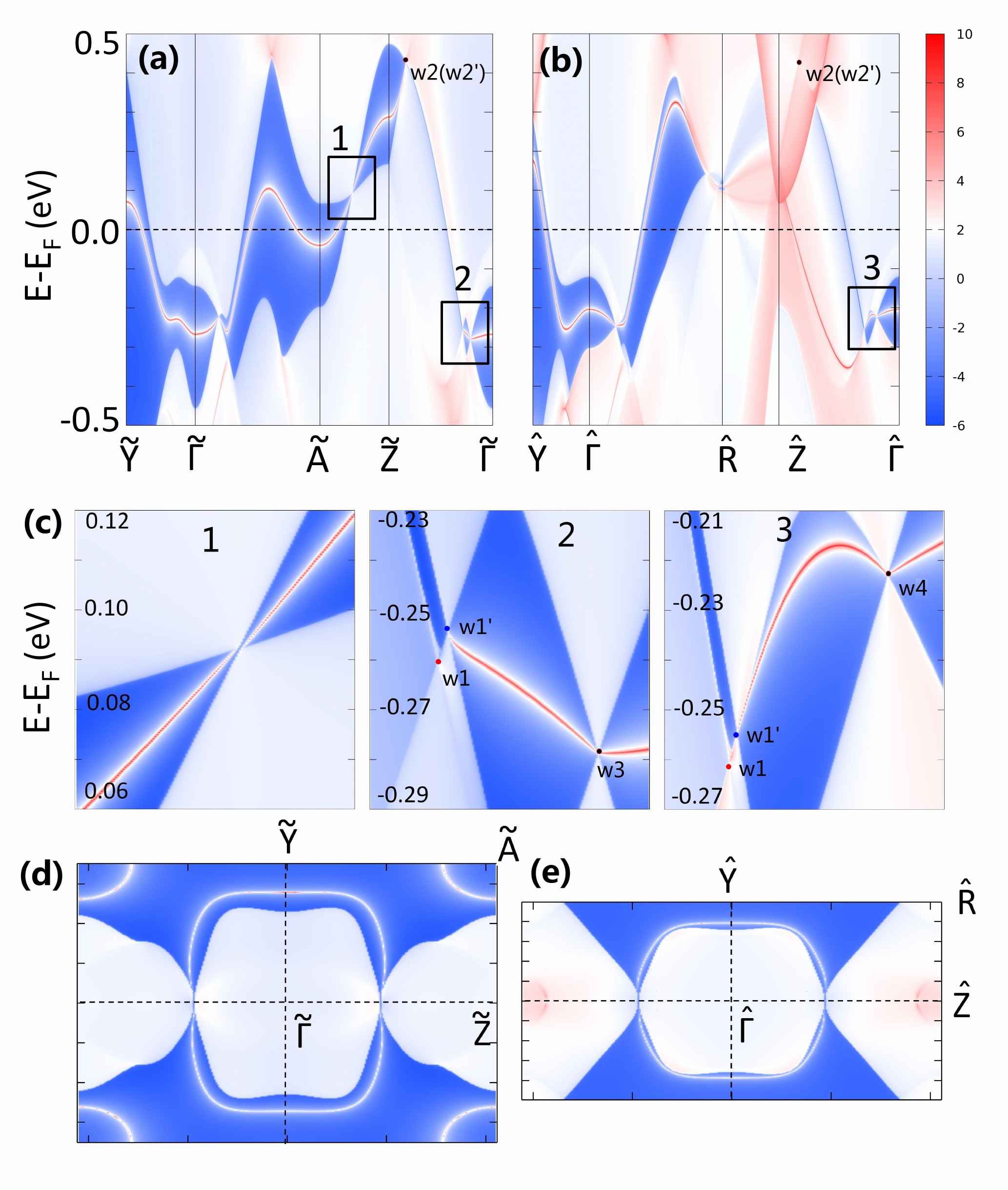}
	\caption{Surface states and Fermi arcs of $r$-CrO$_2$. The energy bands projected on the (a) (100) and (b) (110) surfaces. (c) The regions denoted by 1, 2, 3 in (a) and (b) are enlarged in left, middle, and right panels, respectively. The red and blue dots indicate the projections of WPs with Cherm numbers +1 and -1, respectively. The black dots are projected from two WPs with opposite chirality. The Fermi arcs projected on the (d) (100) and (e) (110) surfaces.
\label{surf-band}}
\end{figure}

One hallmark of Weyl fermions is the existence of nontrivial surface states and Fermi arcs. Based on the maximally localized Wannier functions on the basis of DFT calculations \cite{Marzari2012,Mostofi2008}, we obtain the surface states using Wannier tight-binding Hamiltonian with the iterative Green's function method \cite{Sancho1984} as implemented in WannierTools package \cite{Wu2017wanntool}. To extract the nontrivial surface states of $r$-CrO$_2$, we calculate the surface states projected on (001), (100), and (110) surfaces, respectively. We can find that the topologically protected surface states projected onto (100) and (110) surfaces are clearly visible shown in Figs. \ref{surf-band}(a) and \ref{surf-band}(b), respectively. The surface states on the (001) surface are mostly hidden in the projection of bulk states, which are supplied in SM \cite{SM}. The band gap arising from the type-II WP without SOC is nontrivial due to the band inversion, exhibiting visible surface states along $\tilde{\mathrm{A}}$-$\tilde{\mathrm{Z}}$ [see left panel of Fig. \ref{surf-band}(c)]. The WPs $W1$ ($W1'$) and  $W2$ ($W2'$) projected onto (100) and (110) surfaces are different. However, only $W1'$ is apparent and its projection acts as the termination of surface Fermi arcs, while the projections of other WPs are hidden in bulk states, as shown in Figs. \ref{surf-band}(a)-(c). The WP $W3$ projected on the (100) surface locates on $\tilde{\Gamma}$-$\tilde{\mathrm{Z}}$ [see middle panel of Fig. \ref{surf-band}(c)]. We can see that there are two topological surface states terminated at the projection of $W3$, since this surface Dirac point projected from two $W3$ WPs with opposite chirality. Similarly, the projection of $W4$ on the (110) surface also exhibits two nontrivial surface states which are distributed on two sides of the surface Dirac core, shown in the right panel of Fig. \ref{surf-band}(c). The projection of the bulk Fermi surface onto the (100) and (110) surfaces are shown in Figs. \ref{surf-band}(d) and \ref{surf-band}(e), respectively. The nontrivial Fermi arcs are clearly visible, which makes them can be achieved in angle-resolved photoemission spectroscopy (ARPES) measurements.

\begin{figure}
\setlength{\belowcaptionskip}{-0.4cm}
\setlength{\abovecaptionskip}{-0.1cm}
	\centering
	\includegraphics[scale=0.3]{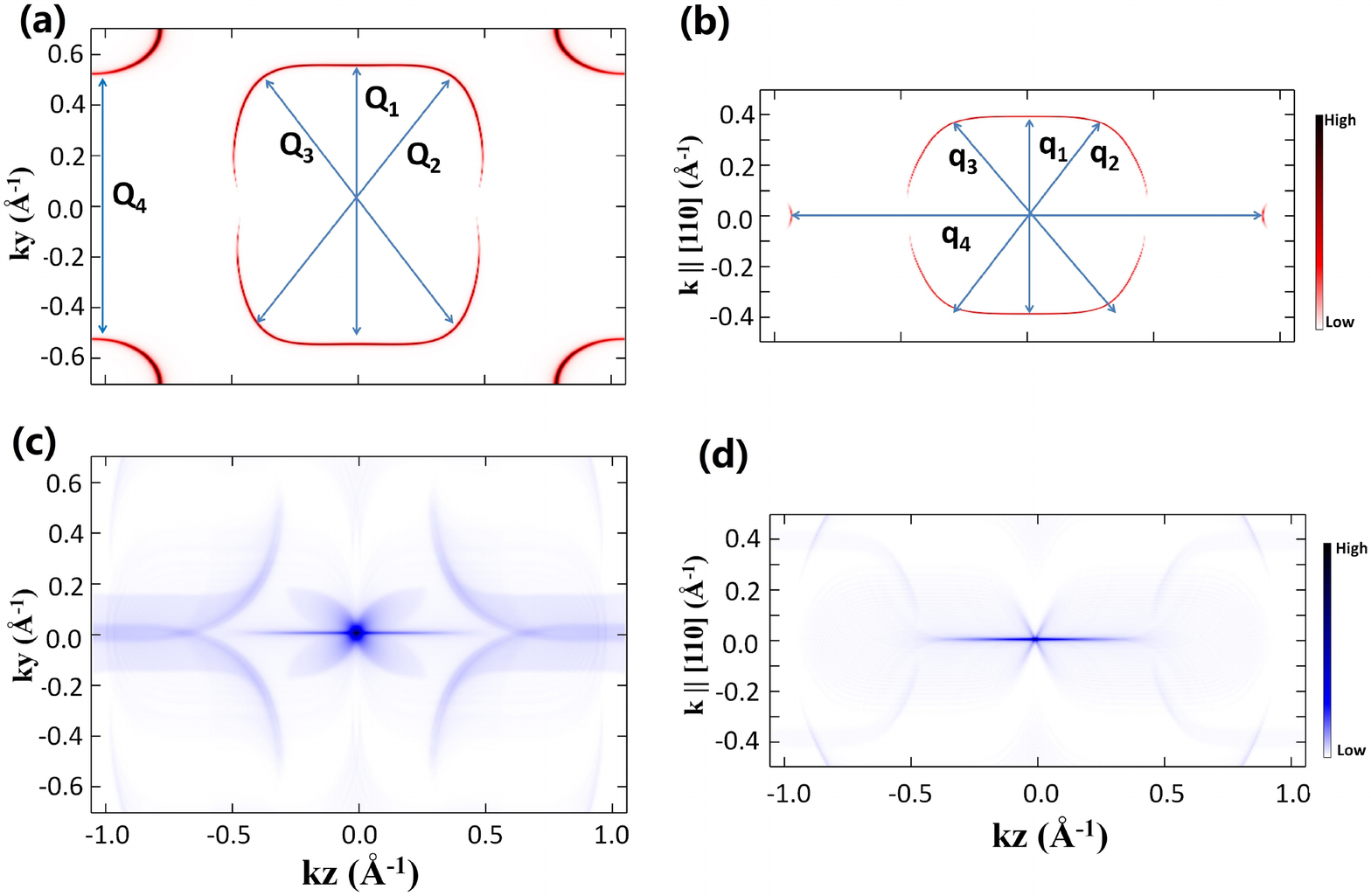}
	\caption{The weighted Fermi arcs and JDOS of $r$-CrO$_2$. The weighted Fermi arcs with only surface contribution on (a) (100) and (b) (110) surfaces. The scattering wave vectors are denoted by $Q_1$ ($q_1$) to $Q_4$ ($q_4$) for the (100) [(110)] surface. The JDOS patterns for (c) (100) and (d) (110) surfaces related to the scattering wave vectors are visible.
\label{QPI}}
\end{figure}

The surface Fermi arcs are crucial properties associated with the topological nontrivial feature of bulk WPs. On the one hand, the surface Fermi arcs can be directly confirmed by ARPES; on the other hand, the nontrivial Fermi arcs pertain solely to a surface, so they will exhibit unique signatures in quasiparticle interference (QPI) and can be observed via spectroscopy measurements using scanning tunneling microscopy (STM) \cite{Inoue1184}. Comparing with the projected Fermi surface of nonmagnetic WSMs in which Fermi arcs are spin textured and time-reversal symmetry prevents scattering between the states with opposite spins, the spin-polarized Fermi arcs in half-metallic CrO$_2$ will exhibit the single spin scattering. Hence, the QPI pattern of surface states is mainly dominated by initial and final states  in momentum space. It can be approximated by the joint density of states (JDOS) as a function of momentum difference $\mathbf{q}$ \cite{PhysRevB.93.041109}
\begin{equation}
J(\mathbf{q}, \epsilon)=\int{d^2 \mathbf{k}_{\parallel}}\rho^{0}(\mathbf{k}_{\parallel},\epsilon)\rho^{0}(\mathbf{k}_{\parallel}+\mathbf{q}, \epsilon),
\end{equation}
where $\rho^{0}(\mathbf{k}_{\parallel},\epsilon)=-\frac{1}{\pi} \mathrm{Im} [\mathrm{Tr} \mathbf{G}(\mathbf{k}_{\parallel},\epsilon)]$ is the momentum-resolved surface density of states of a clean sample, and $\mathbf{G}(\mathbf{k}_{\parallel},\epsilon)]$ is surface Green's function at surface momentum vector $\mathbf{k}_{\parallel}$ and energy $\epsilon$ relative to the Fermi level. The weighted Fermi arcs of $r$-CrO$_2$ with only the surface contribution on the (100) and (110) surfaces are shown in Figs. \ref{QPI}(a) and \ref{QPI}(b). We can see that the open Fermi arcs have a nearly constant spectral density. Based on the curvature of the weighted Fermi arcs, we can identify four different groups of scattering wave vectors for two surfaces, i.e., $Q_1$ to $Q_4$ for the (100) surface and $q_1$ to $q_4$ for the (110) surface. Figs. \ref{QPI} (c) and \ref{QPI}(d) illustrate the JDOS patterns for (100) and (110) surfaces, respectively. The shapes of QPI originate from the corresponding scattering wave vectors are clearly visible. In comparison with the spin-dependent scattering probability in nonmagnetic WSMs, the JDOS in ferromagnetic WSMs are only dependent on momentum difference and are more favorably to be measured  experimentally.

In summary, we propose that FM Weyl fermions with time-reversal breaking can exist in $r$-CrO$_2$ and $o$-CrO$_2$. Based on topology and symmetry analysis, $r$-CrO$_2$ is found to possess TPs in the absence of SOC. By taking into account of SOC, the magnetic group causes a TP splitting into two WPs with opposite chirality, which are located on the magnetization axis respect to the four-fold rotational symmetry. In comparison with $r$-CrO$_2$, $o$-CrO$_2$ with the lower symmetry can not host the TPs. However, the symmetry protected WPs of $o$-CrO$_2$ also leads to alike topological features due to the similar crystal structure and chemical bonds. Impressively, the topologically protected surface states, Fermi arcs, and the related QPI pattern are clearly visible. Our findings strongly suggest that CrO$_{2}$ with FM Weyl fermions have potential topology-related applications at room temperature.

 This work is supported by the National Natural Science Foundation of China (NSFC, Grant Nos.11674148 and 11334004) and the Guangdong Natural Science Funds for Distinguished Young Scholars (No. 2017B030306008).\\

 R.W., Y.J.J. and L.D. contributed equally to this work.

\bibliographystyle{apsrev}

\newpage
\appendix

\setcounter{figure}{0}
\makeatletter

\makeatother
\renewcommand{\thefigure}{S\arabic{figure}}
\renewcommand{\thetable}{S\Roman{table}}
\renewcommand{\theequation}{S\arabic{equation}}
\begin{center}
	\textbf{
		\large{Supplemental Material for}}
	\vspace{0.2cm}
	
	\textbf{
		\large{
			``Room-Temperature Ferromagnetic Topological Phase in CrO$_{2}$: From Tripe Fermions to Weyl Fermions" }
	}
\end{center}

\vspace{-0.2cm}

In this Supplemental Material, we provide topological surface states $r$-CrO$_2$ projected on (001) surface with [001] magnetization, topological features $r$-CrO$_2$ of magnetization along [100] direction, and the symmetry analysis on band topology. Finally, the topological features of $o$-CrO$_2$ are included.



\ \ \\
{\large{\noindent{\textbf{Computational Methods}}}}\\

We perform first-principles calculations based on the density functional theory (DFT) \cite{KohnSI} as implemented Vienna \textit{ab initio} Simulation Package (VASP) \cite{Kresse2SI,KressecomSI}. The core-valence interactions are treated by pseudopotential method with the projector augmented wave (PAW) \cite{BlochlSI,Kresse4SI,Ceperley1980SI} method.  To describe the exchange-correlation potential, we chose the generalized gradient approximation (GGA) with the Perdew-Burke-Ernzerhof (PBE) formalism \cite{Perdew1SI,Perdew2SI}. A plane-wave-basis set with kinetic-energy cutoff of 600 eV is used to get a fully converged electronic structure including spin-orbital coupling (SOC). The Brillouin zone(BZ) is sampled by $21 \times 21 \times 29$ Monkhorst-Pack grid \cite{MonkhorstSI} in self-consistent calculations. In order to account for the magnetic character of the system we performed spin polarized calculations, with the full crystal symmetry implemented per spin species. Since the $3d$-electrons of Cr are strongly correlated, we employ GGA+U scheme \cite{Liechtenstein1995SI} and introduce on-site Coulomb repulsion of U = 3 eV, which is a good agreement with the previous electronic band structures and half-metallic properties of rutile CrO$_2$ \cite{PhysRevLett.80.4305SI}.  In addition, we also verify the topological features with U in the range from 1.5 eV to 6.0 eV, and the similar results are obtained.  The results for the electronic band structures and topological features are also confirmed versus the all-electron calculations in WIEN2K code \cite{SCHWARZ200271SI}.

\ \ \\
{\large{\noindent{\textbf{Topological surface states of $r$-CrO$_2$ projected on (001) surface with [001] magnetization}}}}\\

\begin{figure}
	\centering
	\includegraphics[scale=0.28]{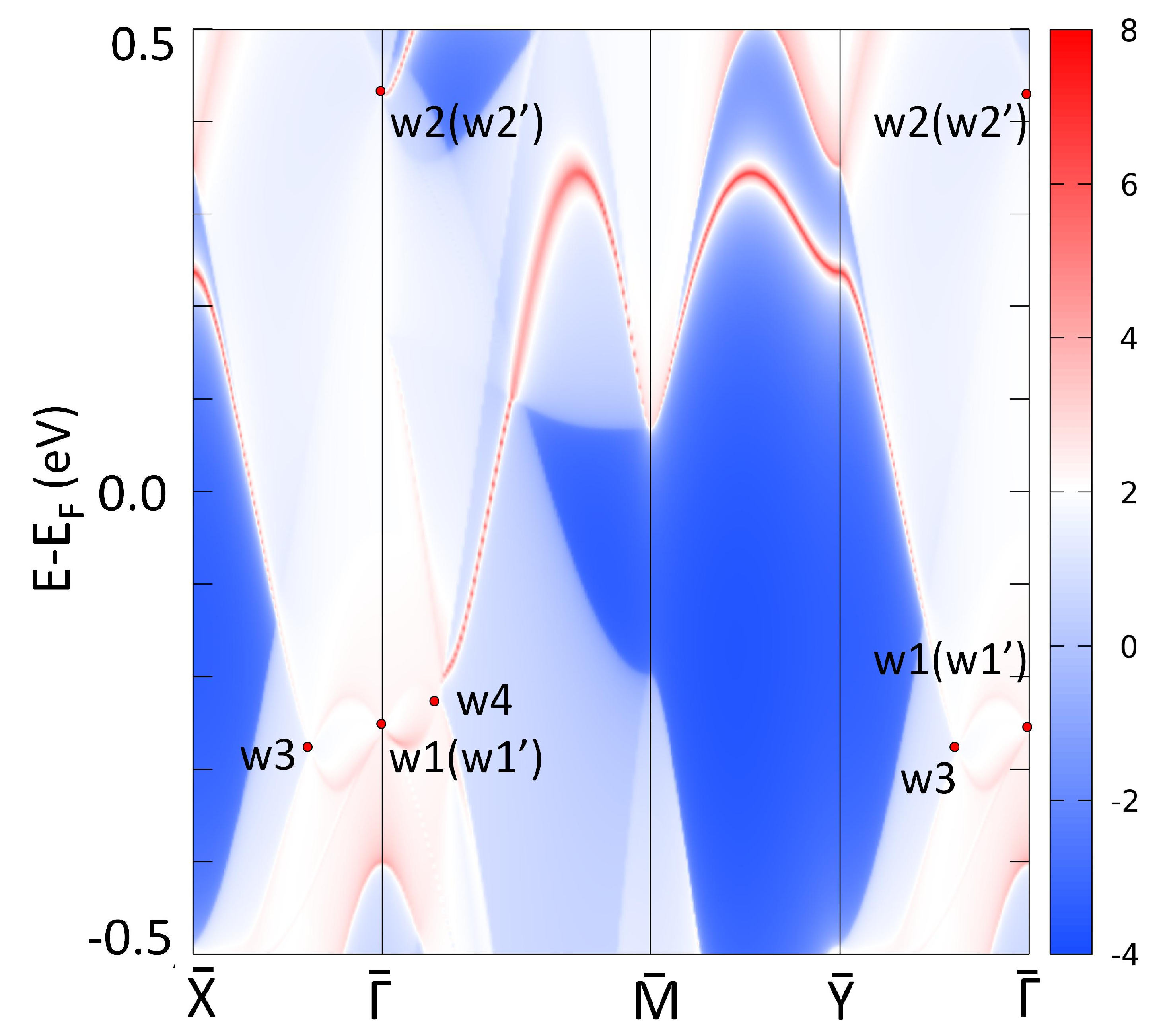}
	\caption{The surface band structures of $r$-CrO$_2$ projected on (001) surface  with [001] magnetization.
  \label{figS1}}
\end{figure}

Unlike the surface states projected on (100) and (110) surfaces, the projections of bulk bands on (001) surface would hide the Weyl nodes and most of topological surface states. However, the WPs projected on this plane are still visible. As shown in Fig. \ref{figS1}, we can clearly see that the projections of Dirac cones are distinct from trivial bulk states.

\ \ \\
{\large{\noindent{\textbf{Topological features $r$-CrO$_2$ of magnetization along [100] direction}}}}\\

Our first-principles calculations also suggest that there are only tiny energy differences among all magnetic configurations in  $r$-CrO$_2$,  meaning that the magnetization directions can be easily manipulated by applied magnetic field. In the presence of SOC, the topological features strongly depend on the magnetic axis, which determines the magnetic space group of system.  Therefore, besides the case of [001] magnetization in the main text, we  also perform the calculations for the magnetization along [100] direction. When the ferromagnetic magnetization is parallel to [100] direction, the system reduces to magnetic space group $D_{2h}(C_{2h})$. As shown in Fig. \ref{figS3}, we can see that the hybridization splits the double-fold degenerate band and there are no crossing nodes along $\Gamma$-Z, since the antiunitary symmetry $\mathcal{T}C_{2z}$ don't allow the existence of WPs on this line. The little group of momentum points (0, 0, $k_z$) along A-Z belongs to $C_1$, so the type-II WP is opened by a gap $\sim$ 5 meV as shown in Fig. \ref{figS3}. In the $k_z=\pi /c$, four other type-II WPs $P1$ are found slightly away from the A-Z axis. The presence of WPs in this plane is allowed by the antiunitary symmetry $\mathcal{T}C_{2z}$. Besides, antiunitary symmetry $\mathcal{T}C_{2y}$ also allow the existence of WPs in $k_x$-$k_z$ plane with $k_y=0$, and two types of WPs $P2$ and $P3$ in this plane are found. We conclude that the independent WPs with magnetization along [100] direction in Table , such as the precise coordinates in momentum space, Chern numbers, and the energies related to the Fermi level $E_F$.

\begin{table}
\caption{Weyl points of $r$-CrO$_2$ with magnetization along [100] direction.  The positions in momentum space, Chern numbers, and the energies relative to $E_F$ are listed, respectively.
The coordinates of the other WPs are related to the ones listed by the symmetries, $\mathcal{I}$, $\mathcal{T}C_{2y}$, and $\mathcal{T}C_{2z}$. The WP $P1$ belongs to type-II.}
  \begin{tabular}{p{0.9 cm}|*{1}{p{3.5cm}} *{3}{p{1.2cm}} }
  \hline
  \hline
    Nodal  & \centering Coordinates [$k_x(2\pi/a)$,    &\centering  Chern & $E-E_F$ \\
    points & \centering $k_y(2\pi/a)$, $k_z(2\pi/c)$] &\centering number & (eV) \\
  \hline
    \centering  $P1$  &\centering (0.2671,    0.2669,   0.5)   &\centering $+1$  &{\centering 0.092} \\
    \centering  $P2$ &\centering  (0.0143,    0,    0.3577)   &\centering $-1$  &{\centering  0.344} \\
    \centering  $P3$ &\centering  (0.1902,    0,   0.1069)   &\centering $+1$  &{\centering -0.279}  \\
  \hline
  \hline
  \end{tabular}
  \label{tables1}
\end{table}

\begin{figure}
	\centering
    \includegraphics[scale=0.3]{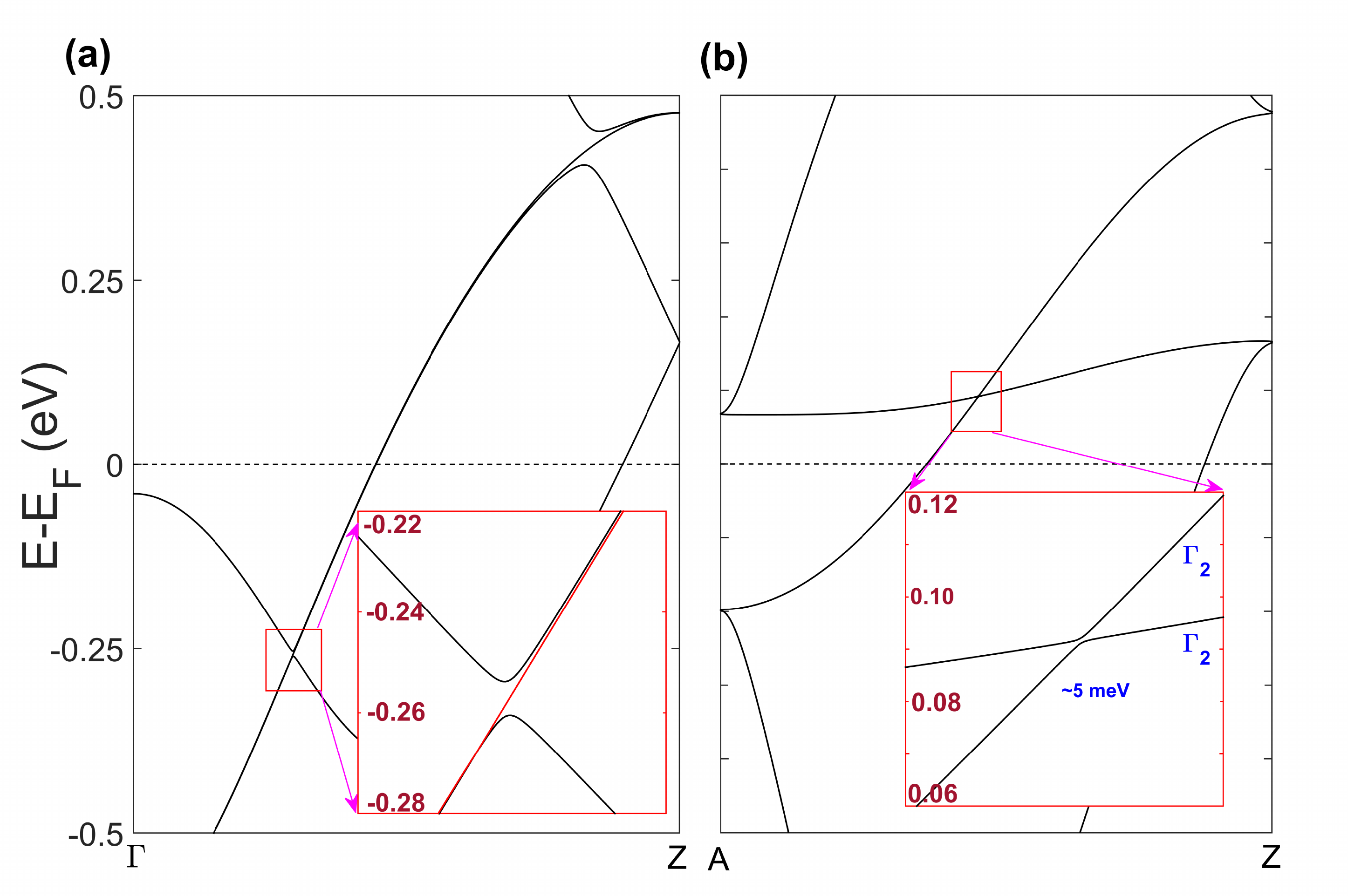}
	\includegraphics[scale=0.03]{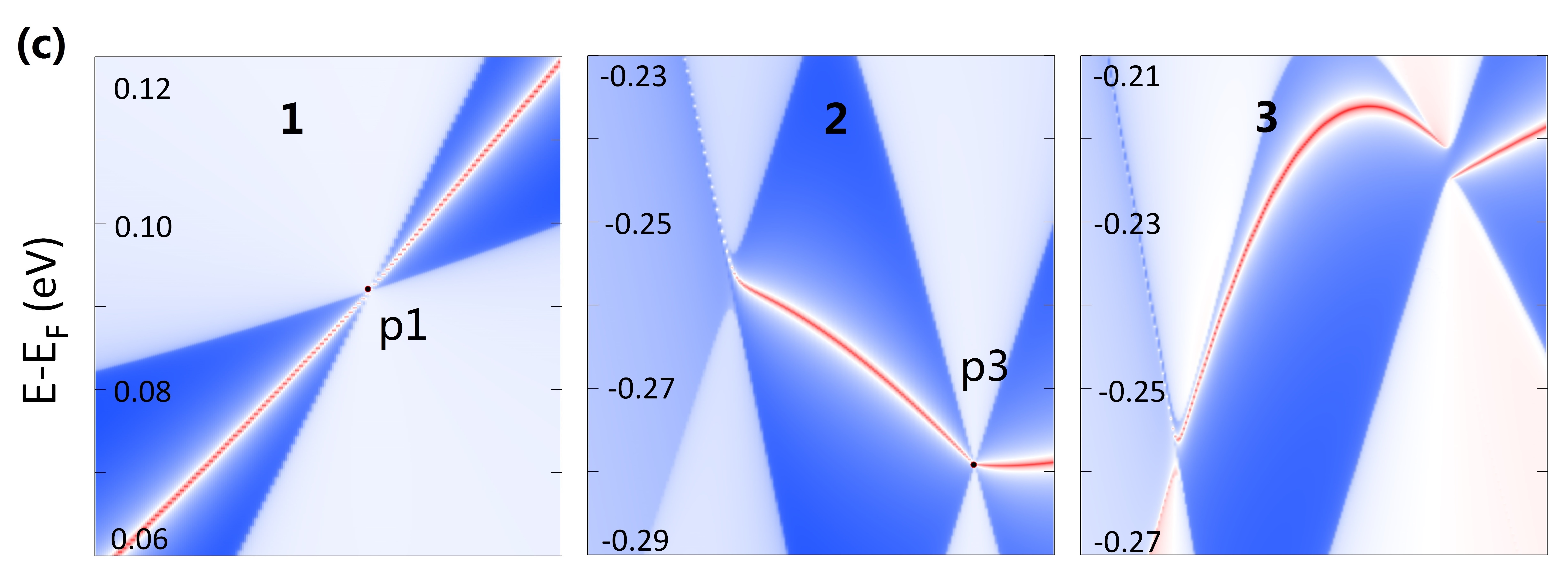}
	\caption{With magnetization along [100] direction, the SOC band structures of $r$-CrO$_2$ along (a) $\Gamma$-Z and (b) A-Z high symmetry lines. The insets show the enlarged view of red squares. (c) The topological surface states. Panels 1, 2, and 3  correspond to the same regions of Fig. 3 in the main text.
  \label{figS2}}
\end{figure}

\begin{figure}
	\centering
	\includegraphics[scale=0.34]{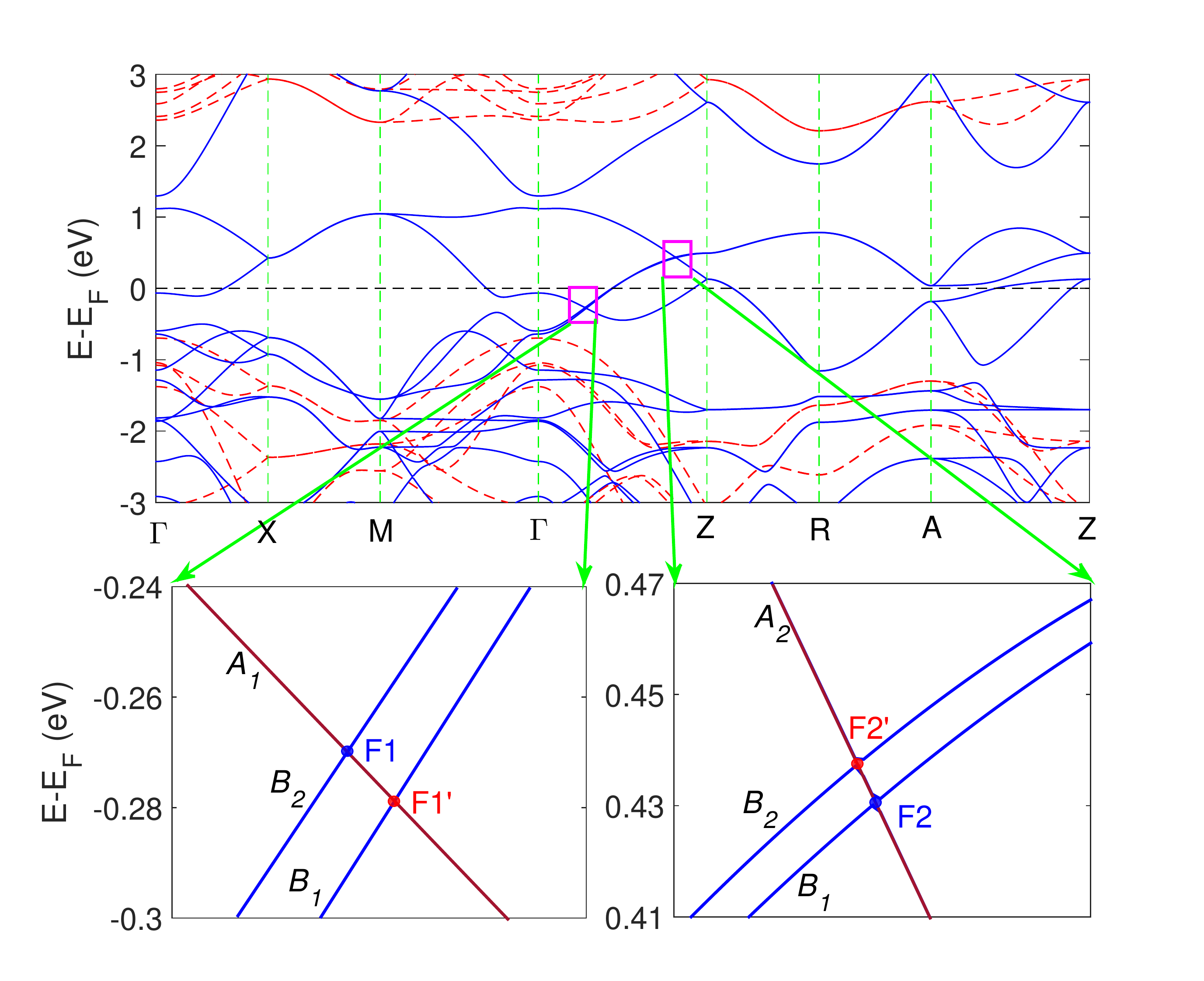}
	\caption{Electronic band structures of $o$-CrO$_{2}$ without SOC. The majority and minority spin bands are denoted by the solid (blue) and dashed (red) lines.  The lower left and right panels  show the enlarged views of two crossing regions along $\Gamma$-Z, respectively. The WPs $F1$ ($F1'$) and $F2$ ($F2'$) are denoted by solid circles, which are marked by the red and blue colors representing the WPs with Cherm numbers +1 and -1, respectively.
  \label{figS3}}
\end{figure}

Due to the weak SOC effects of both Cr and O elements, the surface band structures with magnetization along [100] direction would exhibit similar behaviors in comparison with those with magnetization along [001] direction. However, the different distributions of WPs would essentially possess the different topological surface states. As shown in Fig. \ref{figS2}(c), the panels $1$, $2$, and $3$ correspond to the same regions of Fig. 3 in the main text. We can see that the nontrivial surface states are visible. Two type-II WPs $P1$ with opposite chirality projected onto $\tilde{\mathrm{R}}$-$\tilde{\mathrm{Z}}$ form two nontrivial surface states. Similarly, the projection of $P3$ and the related surface states is located on $\tilde{\mathrm{Z}}$-$\tilde{\mathrm{\Gamma}}$.

\ \ \\
{\large{\noindent{\textbf{The triply-degenerate points of $r$-CrO$_2$ in the absence of spin-orbital coupling}}}}\\

In the absence of SOC, the space group of $r$-CrO$_2$ is $P4_2/mnm$, which contains symmorphic group elements: inversion $\mathcal{I}$, four-fold rotational symmetry $C_4$ axis along [001](or $z$-axis), and and mirror reflection $M_z$. In addition, the space group $P4_2/mnm$ also possesses the nonsymmorphic symmetries as $N_x=\{M_x | \mathbf{\tau}\}$ and $N_{4z}=\{C_{4z} | \mathbf{\tau}\}$, where $\mathbf{\tau}=(\frac{1}{2},\frac{1}{2},\frac{1}{2})$ is the translation vector of one-half of a body diagonal. In real space $(x, y, z)$, we consider the coordinate transformation under $C_{4z}$ and $N_{x}$, and we have
\begin{equation}
\begin{split}
(x, y, z)& \underrightarrow{\ \ \ N_x \ \ \ } (-x+\frac{1}{2}, y+\frac{1}{2}, z+\frac{1}{2})\\
         & \underrightarrow{\ \ \ C_{4z} \ \ \ } (-y-\frac{1}{2}, -x+\frac{1}{2}, z+\frac{1}{2}),\\
(x, y, z)& \underrightarrow{\ \ \ C_{4z} \ \ \ } (-y, x,  z)\\
         & \underrightarrow{\ \ \ N_{x} \ \ \ } (y+\frac{1}{2}, x+\frac{1}{2}, z+\frac{1}{2}).
\end{split}
\end{equation}
It is easy to obtained
\begin{equation}
N_x C_{4z} = C_{2z}T_{(0,-1,0)} C_{4z}N_x,
\end{equation}
where $T_{(0,-1,0)}$ is the translation operator acting on the Bloch wave function, i.e.,  $[N_x, C_{4z}]\neq 0$. In addition, every momentum point on the $\Gamma$-Z is invariant under the $C_{4z}$ and the product $\mathcal{P}M_z$. In momentum space, the little group along $\Gamma$-Z is  $C_{4v}$, which can maintain that there is always a double-fold degenerate band along this line. Hence, two single-fold degenerate bands crossing this double-fold degenerate band will form two TPs denoted as $TP1$ and $TP2$.

\ \ \\
{\large{\noindent{\textbf{Little group of $r$-CrO$_2$ of the $k$-space Hamiltonian in $k_x$-$k_z$ and $k_y$-$k_z$ planes  with [001] magnetization}}}}\\

Including SOC, with [001] magnetization, the little groups of $r$-CrO$_2$ in $k_x = 0$ and $k_y=0$ planes are $\mathcal{T}C_{2x}$ and $\mathcal{T}C_{2y}$, respectively. The general two-band Hamiltonian is
\begin{equation}\label{H}
H(k_x,k_y,k_z)=\sum_{i=x,y,z}d_{i}(k_x,k_y,k_z)\sigma_{i},
\end{equation}
where $d_{i}(k_x,k_y,k_z)$ are real functions and $(k_x,k_y,k_z)$ is momentum vector. For the $k_y$-$k_z$ plane of $k_x = 0$, we have
\begin{equation}
[H(0,k_y,k_z),\mathcal{T}C_{2x}]=0,
\end{equation}
which gives
\begin{equation}\label{H}
H(0,k_y,k_z)=d_y(0,k_y,k_z)\sigma_y+d_z(0,k_y,k_z)\sigma_z.
\end{equation}
The Hamiltonian contains two parameters, two mo-
menta, and WPs can generically live on this high
symmetry plane. Similarly, the little group $\mathcal{T}C_{2y}$ can allow WPs locating on the $k_x$-$k_z$ plane of $k_y=0$.

\ \ \\
{\large{\noindent{\textbf{Symmetries and Weyl Points of $r$-CrO$_2$ with [100] magnetization}}}}\\

When the ferromagnetic magnetization parallel to the [100] direction, the magnetic space group is $D_{2h} (C_{2h})$ and  the following symmetry group remains: $\mathcal{I}$, $C_{2x}$, $\mathcal{T}C_{2y}$, and $\mathcal{T}C_{2z}$.
Along $\Gamma$-Z, antiunitary symmetry $\mathcal{T}C_{2z}$ gives
\begin{equation}
[H(0,0,k_z), \mathcal{T}C_{2z}]=0,
\end{equation}
which implies
 \begin{equation}
d_z(0, 0, k_z)=0.
\end{equation}
Now the Hamiltonian in $kz$ axis is
\begin{equation}\label{H}
H(0,0,k_z)=d_x(0,0,k_z)\sigma_x+d_y(0,0,k_z)\sigma_y.
\end{equation}
The Hamiltonian contains two parameters $d_x$ and $d_y$, one momentum $k_z$, giving rise generically to avoided crossings. No crossing points are possible on $\Gamma$-Z line.
Using the similar analysis , $\mathcal{T}C_{2y}$ and $\mathcal{T}C_{2z}$ can allow WPs on $k_y=0$ and $k_z=\frac{\pi}{c}$ planes, respectively.

\ \ \\
{\large{\noindent{\textbf{Topological features and Weyl points in $o$-CrO$_{2}$}}}}\\

Groundstate CrO$_2$ ($r$-CrO$_{2}$), which crystallizes in the tetragonal structure of the rutile type, is a well-known material because of its half-metallic nature. Under pressure,  crystalline CrO$_{2}$ is known to undergo
the structural transition from  $r$-CrO$_{2}$ to the orthorhombic CrO$_2$ ($o$-CrO$_2$) of the CaCl$_2$ type  at $P = $$12\sim17$ GPa \cite{PhysRevB.73.144111SI}. For the high pressure phase of $o$-CrO$_2$ with space group $Pnnm$ (No. 58), the orthorhombic lattice structures is very similar to the tetragonal unit cell as shown Fig. 1 in the main text. The lattice constants are $a=4.502$ {\AA}, $b=4.429$ {\AA}, and $c=2.968$, which show good agreements with the experimental values \cite{PhysRevB.73.144111SI}.

\begin{figure}
	\centering
	\includegraphics[scale=0.34]{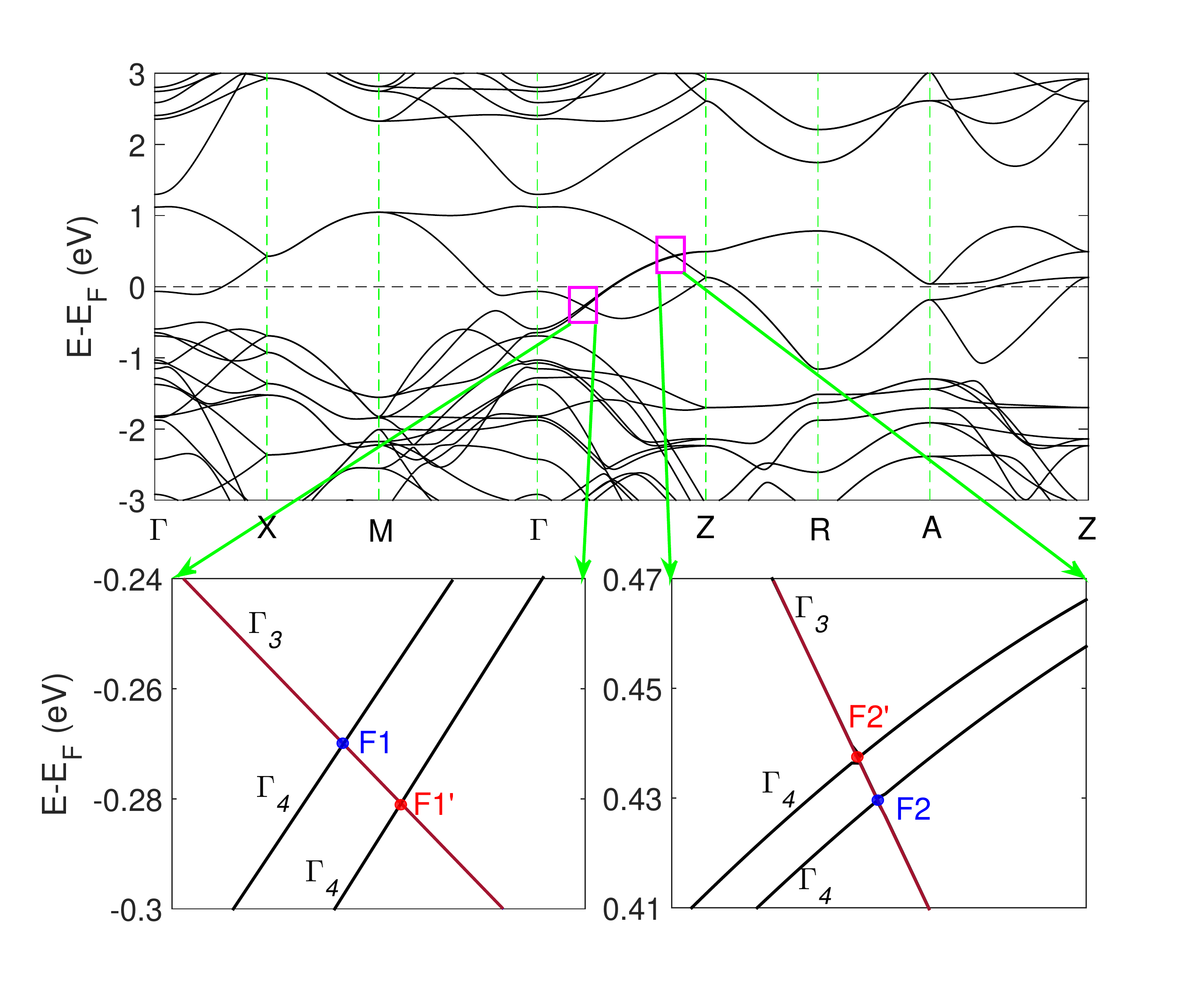}
	\caption{ The SOC electronic band structures of $o$-CrO$_{2}$ with magnetization along [001] direction.  The lower left and right panels  show the enlarged views of two crossing regions along $\Gamma$-Z, respectively. The WPs $F1$ ($F1'$) and $F2$ ($F2'$) are denoted by solid circles, which are marked by the red and blue colors representing the WPs with Cherm numbers +1 and -1, respectively.
  \label{figS4}}
\end{figure}

The spin-polarized electronic band structures of $o$-CrO$_{2}$ without SOC are shown in Fig. \ref{figS3}. We can see that $o$-CrO$_{2}$ also exhibits the half-metallic features and hosts the very similar band structures with $r$-CrO$_2$ due to their alike crystal structure and chemical bonds. However, the detailed band topologies protected by symmetry between $r$-CrO$_2$ and $o$-CrO$_2$  are different. In the absence of SOC, $o$-CrO$_2$ possess the point group $D_{2h}$. The little group along $\Gamma$-Z is $C_{2v}$, which can not protect the TPs on this line but can allow WPs exist. As shown in the lower left and right panels of Fig. \ref{figS3}, there are four symmetry-protected WPs
 formed by the eigenvalues of $C_{2v}$, denoted by $F_1 (F_1 ')$ and $F_2 (F_2 ')$, respectively. In addition, comparison with $r$-CrO$_{2}$, $o$-CrO$_{2}$ doesn't possess WPs along Z-A due to the absence of two-fold rotational symmetry $C_2$ on this line.  In the presence of SOC, the topological band structures are determined by magnetic axis. Here, we only discuss the case of magnetization along [001] direction, and the electronic band structures are shown in Fig. \ref{figS4}. We can see that the SOC little influences on the bands due to the weak SOC strength. With [001] magnetization, the symmetry of o-CrO$_{2}$ reduces to the magnetic double group $D_{2h}(C_{2h})$. The four WPs along $\Gamma$-Z survive, protected by $C_{2z}$. Namely, the crossing bands belong to the eigenvalues $\Gamma_3$ and $\Gamma_4$ of $C_{2z}$. The antiunitary symmetries $\mathcal{T}C_{2x}$ and $\mathcal{T}C_{2y}$ allow existence of WPs in high-symmetry planes $k_x$-$k_z$ and $k_y$-$k_z$, and the related WPs $F3$ and $F4$ are found. In addition, we also find that another kind of WPs $F5$ related by $\mathcal{P}$ and $C_{2z}$. The WPs $F5$ are generic nodal points and not stable in principle. We conclude that the WPs of $o$-CrO$_2$ with magnetization along [001] direction in Table \ref{tables2}, such as the precise coordinates in momentum space, Chern numbers, and the energies related to the Fermi level $E_F$.

\begin{table}
\caption{Weyl points of $o$-CrO$_2$ with magnetization along [001] direction.  The positions in momentum space, Chern numbers, and the energies relative to $E_F$ are listed, respectively.
The coordinates of the other WPs are related to the ones listed by the symmetries, $\mathcal{P}$, $\mathcal{T}C_{2y}$, $\mathcal{T}C_{2x}$, and $C_{2z}$. }
  \begin{tabular}{p{0.9 cm}|*{1}{p{3.5cm}} *{3}{p{1.2cm}} }
  \hline
  \hline
    Nodal  & \centering Coordinates [$k_x(2\pi/a)$,    &\centering  Chern & $E-E_F$ \\
    points & \centering $k_y(2\pi/b)$, $k_z(2\pi/c)$] &\centering number & (eV) \\
  \hline
    \centering  $F1$  &\centering (0,   0,    0.1378)   &\centering $-1$  &{\centering -0.270} \\
    \centering  $F1'$ &\centering (0,   0,    0.1397)   &\centering $+1$  &{\centering  -0.281} \\
    \centering  $F2$ &\centering  (0,   0,    0.4089)   &\centering $-1$  &{\centering  0.430}  \\
    \centering  $F2'$ &\centering (0,   0,    0.4078)   &\centering $-1$  &{\centering  0.435}  \\
    \centering  $F3$ &\centering  (0.1891,    0,    0.1013)   &\centering $+1$  &{\centering -0.306}  \\
    \centering  $F4$ &\centering  (0,   0.1443    0.1224)   &\centering $+1$  &{\centering  -0.279}  \\
    \centering  $F5$ &\centering  (0.1643,    0.0429,    0.0991)   &\centering $-1$  &{\centering -0.282} \\
  \hline
  \hline
  \end{tabular}
  \label{tables2}
\end{table}


\end{document}